\documentclass{aa}
\usepackage{graphicx}
\usepackage{txfonts}
\begin{document}

\title{Outflowing disk formation in B[e] supergiants due to rotation and bi--stability in radiation driven winds}


   \author{M. Cur\'{e} \inst{1}
          \and
          D. F. Rial      \inst{2}
          \and
          L. Cidale    \inst{3}
          }

   \offprints{M. Cur\'{e}}

\institute{
Departamento de F\'{\i}sica y Meteorolog\'{\i}a, Facultad de Ciencias, 
Universidad de Valpara\'{\i}so, Valpara\'{\i}so, Chile.
\email{michel.cure@uv.cl}
\and 
Departamento de Matem\'{a}ticas, Facultad de Ciencias Exactas y Naturales, 
Universidad de Buenos Aires, Argentina.
\email{drial@dm.uba.ar}
\and
Facultad de Ciencias Astron\'{o}micas y Geof\'{i}sicas, Universidad Nacional de La Plata, La Plata, Buenos Aires, Argentina.
\email{lydia@fcaglp.unlp.edu.ar}
}

\date{Received November ...,2004, ; Accepted ..., 2005}

\abstract{
The effects of rapid rotation and bi--stability upon the density contrast between the equatorial and polar directions of a B[e] supergiant are re--investigated. Based upon a new slow solution for different high rotational radiation driven winds  (Cur\'e 2004) and the fact that bi--stability allows a change in  the line--force parameters  ($\alpha$, $k$, and $\delta$), the equatorial densities  are about $10^2$--$10^4$ times higher than the polar ones. These values are in qualitative agreement with the observations.

} 
\maketitle
\titlerunning{Outflowing Disk Formation.}
\authorrunning{Cur\'e, Rial & Cildale}

\section{Introduction}

B[e] supergiants belong to a post-main sequence evolutionary stage of massive stars, their characteristics are: Balmer lines in emission, sometimes with P Cygni profiles with equivalent
widths for $H_{\alpha}$ greater than $100\,\AA{}$, low--excitation permitted emission lines,
predominantly of singly ionized metals, forbidden emission lines of [O\, I] and [Fe\, II] and
strong near/mid infrared excess due to hot circumstellar dust, indicating dust temperatures 
of $1000\,$K  (Zickgraf et al.~1986,~1992).

A large percentage of B[e] supergiants ($\sim 70\% - 80\%$) show a hybrid spectrum (Zickgraf 
et al.~1985,~1986) that comes from the simultaneous observations of optical narrow low--excitation 
emission lines  ($\sim100$ $km\, s^{-1}$) and broad UV high--excitation absorption lines 
(C IV, Si IV and N V) with terminal velocities similar to early B supergiants winds, of order 
of $\sim1000$ $km\, s^{-1}$. An empirical model that describes these hybrid spectral 
characteristics was suggested by Zickgraf et al.~(1985) in terms of a two--component stellar wind, 
consisting of a fast radiation--driven wind (Castor et al.~1975, hereafter CAK) from polar 
latitudes and a slow and dense expanding disk in the equatorial regions of the star. This 
two--component wind has been confirmed from medium resolution spectropolarimetry for the evolved 
B[e] star HD 87643 (Oudmaijer et al.~1998).

The bi--stability mechanism induced by rotating radiation driven winds was introduced by Lamers 
\& Pauldrach (1991) to explain the formation of outflowing disks around early--type stars. This 
mechanism, which causes a drastic change in the wind structure, was thought to be related to 
the behaviour of the Lyman continuum optical depth, $\tau_L$, with the stellar latitude. 
For $\tau_L < 1$ the wind is fast but for $\tau_L \gtrsim 3$ the wind is slow. The bi--stability 
manifests itself at certain spectral type or temperature. Lamers et al.~(1995) have determined 
terminal velocities, $V_{\infty}$, of a large sample of early--type stars concluding that the 
bi--stability jump is present around $T_{\mathrm{eff}}=21,000\,$K. Vink et al.~(1999) have 
theoretically shown that the bi--stability mechanism is located at $T_{\mathrm{eff}}=25,000\,$K 
and that it is mainly due to the radiative acceleration by iron, caused by the recombination of 
$Fe$ $IV$ to  $Fe$ $III$.

Pelupessy et al.~(2000) have calculated density contrast (ratio between equatorial and polar 
densities) in a B[e] supergiant for rotationally induced bi-stability models applying  
multi--scattering line-force parameters above and below the critical temperature of the 
bi--stability jump. They show that the ratio between equatorial and polar densities is about 
$\sim 10$, and state that this value is a factor $10$ times smaller than Bjorkman's (1998) 
calculations. Porter (2003) also asserts the difficulties in modelling optical-near-IR emission 
for B[e] supergiants with disk model's density structure produced by either a bi-stability wind 
or a Keplerian viscous disk.

Pelupessy et al.~(2000) wind solutions were computed considering values of  
$\Omega = v_{rot}/v_{brkup} \lesssim 0.6 $, where $v_{rot}$ is the equatorial rotational speed 
and $v_{brkup}$ is the break--up speed. However, B[e] supergiants are located near the Eddington 
limit (Zickgraf et al.~1986). Here, critical rotation speed is reached at a much lower 
star's rotational speed. Velocities of about 200 $km\, s^{-1}$ would make the star rotate 
sufficiently close to the break-up speed to produce observable effects. Consequently, the wind 
characteristics near the equator are expected to differ from the polar wind. Langer~(1998) proposes 
the $\Omega$--mechanism for these type of stars. His model calculations suggest that despite the 
loss of angular momentum due to mass loss and increasing radius a rapidly rotating massive 
main-sequence star could remain for a substantial fraction of its lifetime close to the so-called 
$\Omega$--limit which designates the limit of critical (or break--up) rotation. 

In order to investigate the influence of rotation in radiation driven winds (even for a star 
rotating up to break--up rotational speed), Cur\'{e} \& Rial (2004) performed a topological 
analysis of the rotating CAK model, finding that the line--force parameter $\delta$ (Abbott~1982), 
that accounts for changes in the ionization of the wind, leads to a bifurcation in the solution 
topology as a function of the rotational speed, shifting downstream the location of the critical 
point of the CAK (x--type) singular point.  Thus a higher mass--loss rate and lower terminal 
velocity wind is attained in comparison to the frozen--in ionization ($\delta=0$) case. \\
Furthermore, Cur\'{e} (2004) proves that the standard solution (hereafter the fast solution) 
of the m--CAK wind model (Friend \& Abbott 1986, Pauldrach et al.~1986) vanishes for rotational 
speeds of $\sim$0.7 -- 0.8 $v_{brkup}$, and there exists a new solution, that is much denser and 
slower than the known standard  m--CAK solution. We will call it hereafter the {\textit{slow}} 
solution.\\

The purpose of this work is to re-investigate the formation of an equatorial disk--wind for 
rapidly rotating B[e] supergiants, taking into account: 1) the fast and slow solutions of rotating 
radiative driven winds that depend on the assumed rotational speed $\Omega$ and 2) bi--stability 
line--force parameters.\\ 
In section \ref{biparametros} we discuss the adopted line--force parameters. In section \ref{resultados} we show results for CAK and m--CAK models with fast and slow solutions. Section \ref{delta} is devoted to the analysis of the influence of changes in ionization throughout the wind. Discussion and conclusions are presented in sections \ref{diskusion} and \ref{conclu}, 
respectively.

\section{Rotating bi-stability parameters \label{biparametros}}

In order to investigate the influence of the rotation and the bi--stability jump in forming a 
disk--wind, we solve the non--linear momentum equation for the CAK wind and for the m--CAK wind, 
in both polar and equatorial directions. Details and calculation methods here used for the CAK 
wind are found in Cur\'e \& Rial~(2004) and for m--CAK in Cur\'e ~(2004). 

Prior to solving CAK and m--CAK momentum equations, we have to know the line--force parameters.
The parametrization from Abbott~(1982) of the line--force uses 3 parameters 
($\alpha$, $k$ and $\delta$). These parameters are obtained from the fitting of the net 
line acceleration of hundreds of thousands NLTE line transitions computed by solving 
consistently the radiative transfer and hydrodynamics. A different approach for calculating 
the line--force parameters that include multi--line effects for temperatures above and below 
the bi--stability jump was developed by Vink et al.~(1999). They use the Monte--Carlo 
technique and assume a $\beta$--field for the velocity law. 

We want to stress, that no calculation of the line--force parameters has been performed for the 
slow solution and this is beyond the scope of this study. Therefore, in order to solve the wind 
momentum equation, we adopt Pelupessy et al.~(2000) line--force parameters, $\alpha$ and $k$, 
since they have been calculated for both sides of the bi--stability jump. These line--force 
parameters are summarized in Table \ref{tabla1}.
   \begin{table}
      \caption[]{Bi--stability line force parameters}
         \label{tabla1}
     $$ 
         \begin{array}{p{0.1\linewidth}ccccc}
            \hline  \hline
            \noalign{\smallskip}
            T ${[\mathrm{K}]}$ & \alpha & & k & & \delta \\
            \noalign{\smallskip}
            \hline
            \noalign{\smallskip}
            30,000 & 0.65 & &0.06 & & \, 0   \\
	    17,500 & 0.45 & &0.57 & & \, 0   \\
            \noalign{\smallskip}
            \hline
         \end{array}
     $$ 
   \end{table}
\section{Wind model results \label{resultados}}
In order to compare our results with the ones from Pelupessy et al.~(2000), we adopt the 
same B[e] supergiant star they used: $T_{\mathrm{eff}}=25,000\,$K, $M/M_{\sun}=17.5$, 
$L/L_{\sun}=10^5$ and solar abundance. For the lower boundary conditions for polar and 
equatorial directions, we use the same procedure as Stee \& de Araujo ~(1994), i.e., after 
solving the momentum equation (CAK and m--CAK) in the polar direction, with the surface 
boundary condition, $\tau=2/3$ (electron scattering optical depth), we obtain the value of 
the polar photospheric density, $\rho_{p}(R_{\ast}$). This value of the photospheric density 
is then used as surface boundary condition for the equatorial direction, for both CAK and 
m--CAK wind models, respectively.

In this work we have not taken into account neither the change in the shape of the star nor 
gravity darkening (von Zeipel effect) nor the modification of the finite--disk correction 
factor due to the rotation (Cranmer \& Owocki~1995 equation [26], Pelupessy et al.~2000). 
However, we expect that these effects may hava a small influence on the fast solution 
(see section \ref{fast-sol}). The study of their influences in the slow solution will be the 
scope of a forthcoming article.

\subsection{The rotating bi-stability CAK wind}
As a first step to study the combined effect of bi--stability and high rotation speed we 
investigate the CAK wind. We solve the isothermal ($T_{\mathrm{eff}}=25,000$) CAK wind for 
our B[e] supergiant, for the polar(equatorial) direction we use the line--force parameters 
determined for temperatures above(below) the bi--stability jump, see Table \ref{tabla1}.

   \begin{table}
      \caption[]{Parameters of the calculated CAK models: terminal velocity, 
      $V_{\infty}$ ($km\, s^{-1}$), $F_m$ the local mass loss rate ($10^{-6}\,M_{\sun\,} yr^{-1}$) 
      and $r_{c}$, the location of the critical point.}
         \label{tabla2}
     $$ 
         \begin{array}{lccccccc}
            \hline \hline
            \noalign{\smallskip}
            & \Omega  & &V_{\infty} && F_{m} && r_{c}/R_{\ast}\\
            \noalign{\smallskip}
            \hline
            \noalign{\smallskip}
            pole    & 0.0 &\,\,& 826 & &0.181 & &1.60 \\
            \hline
            equator & 0.6 & & 379 && 0.718 & &9.83   \\
            equator & 0.7 & & 348 && 0.730 & &11.47   \\
            equator & 0.8 & & 319 && 0.743 & &13.13   \\    
            equator & 0.9 & & 293 && 0.756 & &14.80   \\
            equator & 0.99& & 273 && 0.770 & &16.33    \\  
            \noalign{\smallskip}
            \hline
         \end{array}
     $$ 
   \end{table}
   \begin{figure}
   \centering
   \includegraphics[width=8cm] {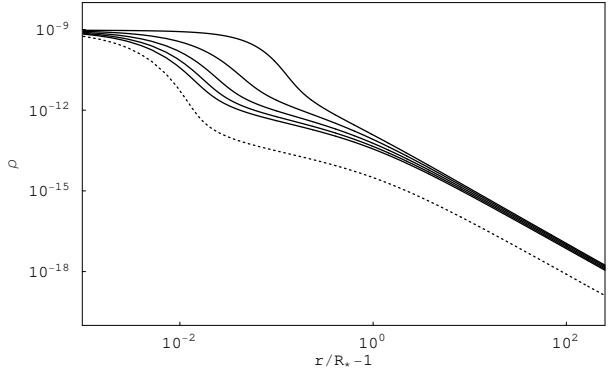}
      \caption{CAK model: density (in $gr\,cm^{-3}$) versus $r/R_{\ast}-1$. Polar density 
      is in dashed--line; equatorial densities are in continuous--line, the higher is $\Omega$, 
      the higher is the density ($\Omega=0.6$, $0.7$, $0.8$, $0.9$, $0.99$).
      }
      \label{fig1}
   \end{figure}
%
   \begin{figure}
   \centering
   \includegraphics[width=8cm] {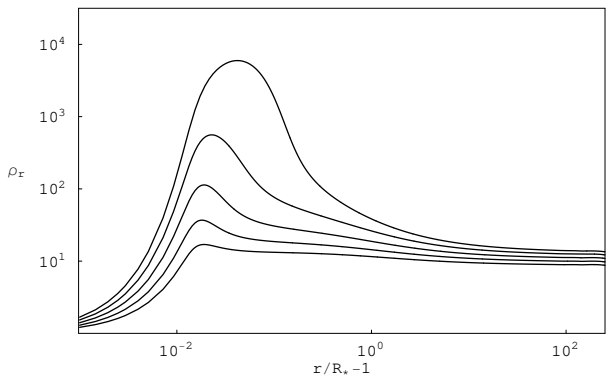}
      \caption{CAK model: density ratio $\rho_{e}(\Omega)/\rho_{p}$ versus $r/R_{\ast}-1$. 
      The higher is $\Omega$, the higher is the density contrast. Curves are  for $\Omega=0.6$, 
      $0.7$, $0.8$, $0.9$, $0.99$, respectively.
      }
      \label{fig2}
   \end{figure}
%
%
Figure \ref{fig1} shows the density profile in both directions, polar (dotted line) 
and equatorial (continuous lines), for different values of the $\Omega$ parameter and 
Figure \ref{fig2} shows ratios between the equatorial, $\rho_{e}$, 
and polar, $\rho_{p}$, densities. 
Our result, without finite--disk correction factor, gives a density contrast similar 
to the one obtained for $\Omega=0.6$ by Pelupessy et al.~(2000), who use the finite 
disk correction factor. Density contrast for $\Omega > 0.7$ are a few times
denser. While the CAK model can be calculated for any rotational speed, Pelupessy et 
al.~(2000) could not calculate m--CAK models for $\Omega > 0.6$,  the reason is because 
the fast solution ceases to exist (Cur\'e~2004).
From Figure \ref{fig2}, we can see that a disk wind structure is formed, the higher is
 $\Omega$, the denser and larger is the disk close to the photosphere and up to 
 $r \sim 2 R_{\ast}$, region where the equatorial density reaches (as function of $\Omega$)
  values of hundreds times the value of the polar density. Far from the stellar surface
   and up to hundreds stellar radii, the ratio $\rho_{e}/\rho_{p}$ reaches values of 
   $\sim 10$.\\
Hence, a rotating bi--stability radiation driven wind forms a disk at the equatorial 
latitudes of these stars. Table \ref{tabla2} summarizes the polar and equatorial values 
of the terminal velocity, $F_m$, the 'local mass loss rate' (see Pelupessy et al.~2000, 
equations [10] and [11]) -- the total mass loss rate if the solution for this latitude 
were valid for a spherical star -- and the location of the critical (singular) point.
The terminal velocities are higher than the observed ones in B[e] supergiants 
disks (Zickgraf,~1998). This fact might be due to the assumed line--force parameters. 
There are no calculations of these parameters for the CAK model with rotational speeds. 
Therefore, these results represent our first approximation to modeling the outflowing 
disks of these 
objects.

\subsection{The rotating bi-stability m--CAK wind \label{m-cak}}
   \begin{figure}
   \centering
   \includegraphics[width=8cm] {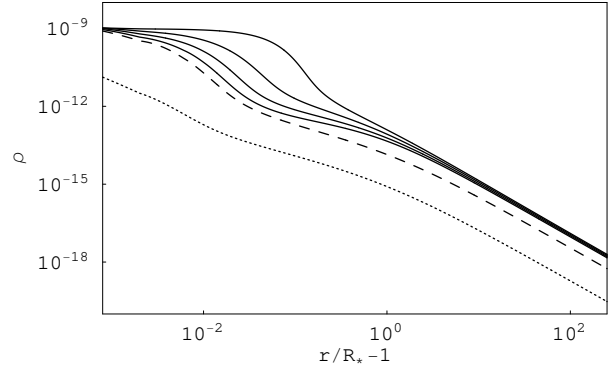}
      \caption{m--CAK model: density (in $gr\,cm^{-3}$) versus $r/R_{\ast}-1$. Polar density 
      is in dotted--line; equatorial density for $\Omega=0.6$ (fast solution) is in dashed--line 
      and equatorial densities  for $\Omega=0.7$, $0.8$, $0.9$, $0.99$ are in continuous--line, 
      the higher is $\Omega$, the higher is the density.
      }
      \label{fig3}
   \end{figure}
   \begin{figure}
   \centering
   \includegraphics[width=8cm] {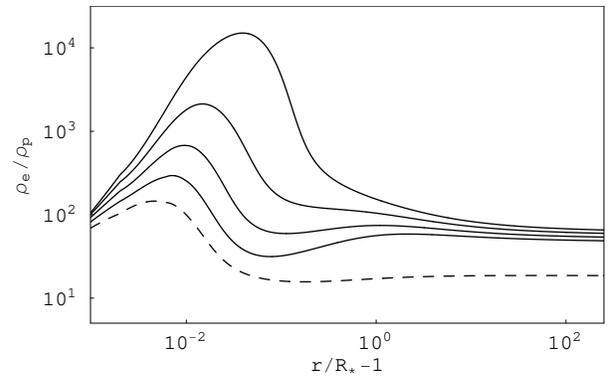}
      \caption{m--CAK model: density contrast versus $r/R_{\ast}-1$, dashed--line is for 
      $\Omega=0.6$ and continuous--line are for $\Omega=0.7$, $0.8$, $0.9$, $0.99$. The higher 
      is $\Omega$, the higher is the density contrast.
      }
      \label{fig4}
   \end{figure}

We present here the results of m--CAK wind models (see solution scheme in Cur\'e~2004), 
taking the same set of line--force parameters given in Table \ref{tabla1} and different 
values of the rotational speed for the equatorial direction. Due to the fact, that the 
existence of the fast or slow solution depends on the rotational speed, we analyse the 
fast solution for $\Omega=0.6$ ($\sim$ upper limit of $\Omega$ for fast solutions) and 
then for higher rotational speeds we obtain slow solutions.  We show that a disk--wind 
with a large density contrast is formed for high rotational speeds. Our results are 
summarized in Table \ref{tabla3}, displayed in figures \ref{fig3} and \ref{fig4} and 
will be discussed below.

\subsubsection{The fast solution \label{fast-sol}}
In order to compare our results with the ones from Pelupessy et al.~(2000), we  solve the 
m--CAK momentum equation, for $\Omega=0.6$. Note that, there are no slow solutions that 
satisfy the lower boundary condition for this value of $\Omega$.\\
This fast solution is shown in figures \ref{fig3} and \ref{fig4} by dashed--line. Its density 
is lower than the densities from slow solutions (continuous--lines, see section 
\ref{slow-sol}) and higher than the polar density ($\Omega=0$). We obtain a density 
contrast of about $10$ for almost all the wind, similiar to Pelupessy et al.~(2000) 
result. Since Pelupessy et al.~(2000) included the effects of: change in the shape of 
the star as a function of the rotational speed, dependence of the temperature on the 
latitude and the finite disk correction factor due to an oblate star, we expect 
that these effects have a small influence on the fast solutions.\\
The increase in the density contrast in the region close to the photosphere 
(see figure  \ref{fig4}) is due to the centrifugal force and the consequently higher 
mass--loss rate of the fast solution when rotation is included (Friend \& Abbott 1986).

\subsubsection{The slow solution \label{slow-sol}}
We have calculated slow solutions from the m--CAK momentum equations in the equatorial 
direction for rotational speeds for $\Omega=0.7$, $0.8$, $0.9$, $0.99$. Density profiles 
from figures \ref{fig1} and \ref{fig3}, for both CAK and m--CAK solutions respectively, 
show a similar behaviour. The differences near the photosphere and up to $\sim 2 R_{\ast}$ 
are due to the finite disk correction factor, $f_D$, but for radii larger than two stellar 
radii, both density structures are almost the same. Near the photosphere $f_D$ is less than 
one, thus a lower mass loss rate is attained. Further out in the wind the value of $f_D$ is 
greater than one, so the plasma is accelerated to higher terminal velocities.
 
Figure \ref{fig4} shows the density contrast profile, which are larger than the 
corresponding CAK models (see figure \ref{fig2}). This is due to the inclusion of 
$f_D$ in the momentum equation, giving a polar flow ($\Omega=0$) less dense than the 
CAK case. Density contrasts reach values around thousand for radii less than 
$\sim 2 R_{\ast}$ and a value of hundred is maintained by the wind up to hundreds of 
stellar radii, almost independently of $\Omega$. This result concerning the disk behaviour 
is in qualitative agreement with the values estimated from observations by Zickgraf 
et al.~(1985,~1986,1992), Zickgraf~(1998), Oudmaijer et al.~(1998) and  Bjorkman~(1998).

   \begin{table}
      \caption[]{Parameters of the calculated m--CAK models: terminal velocity, 
      $V_{\infty}$ ($km\, s^{-1}$), $F_m$, the local mass loss rate
       ($10^{-6}\,M_{\sun} \,yr^{-1}$) and $ r_{c}$ the location of the critical point.}
         \label{tabla3}
     $$ 
         \begin{array}{lccccccc}
            \hline \hline
            \noalign{\smallskip}
            & \Omega  & &V_{\infty} && F_{m} && r_{c}/R_{\ast}\\
            \noalign{\smallskip}
            \hline
            \noalign{\smallskip}
            pole    & 0.0 &\,\,& 2287&& 0.109  & &1.027 \\
 	    \hline
            equator^{\mathrm{a}}  & 0.6 & &  747 && 0.724 && 1.057 \\    
	        equator & 0.7 & &  340 && 0.838 && 12.75 \\
            equator & 0.8 & &  312 && 0.851 && 14.21 \\    
            equator & 0.9 & &  289 && 0.864 && 15.07 \\
            equator & 0.99 & & 266 && 0.875 && 17.14 \\            
	    \noalign{\smallskip}
            \hline
         \end{array}
     $$ 
     \begin{list}{}{}
	\item[$^{\mathrm{a}}$] Fast solution values.
     \end{list}
   \end{table}

\section{Changes in ionization throughout the wind \label{delta}}
\begin{figure}
   \centering
   \includegraphics[width=8cm] {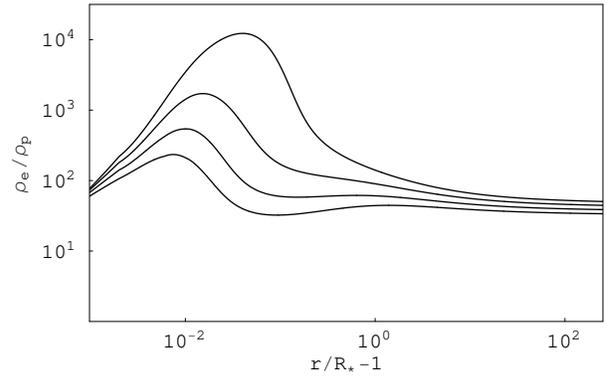}
      \caption{Density contrast for the m-CAK wind with $\delta \ne 0$. 
      Parameters $\alpha$ and $k$ are given in Table \ref{tabla1}.
      }
      \label{fig5}
   \end{figure}

Prinja et al.~(2005) investigate the ionization structure of early-B supergiant winds 
and  demonstrate that the wind ionization increases with distance from the star. This 
structure is different that the one exhibits by an O--star wind. Therefore in order to 
study the effect of changes in ionization with radial distance in a fast rotating 
supergiant, we explore the influence of the parameter $\delta$ in the m--CAK model.\\
We adopt $\delta$ values from Abbott~(1982), for the polar direction, $\delta=0.12$ 
($T=30,000$) and for the equatorial direction, $\delta=0.089$ ($T=20,000$). The 
$\alpha$ and $k$ values are the ones from Table \ref{tabla1}.  Figure \ref{fig5} shows 
density contrast profiles for $\Omega \ge 0.7$, which exhibit the same behaviour as in 
the frozen--in ionization case (section \ref{slow-sol}) and values of the same order. We 
have also calculated m--CAK models with the $\delta$ line--force parameter from Shimada 
et al.~(1994), arriving at the same conclusion, that the $\delta$--parameter has a small 
influence in the density contrast value. Model parameters are summarized in Table 
\ref{tabla4}.

   \begin{table}
      \caption[]{Parameters of the calculated m--CAK models with $\delta \ne 0$ from 
      Abbott (1982), terminal velocity, $V_{\infty}$ ($km\, s^{-1}$), the local mass loss 
      rate, $F_m$ ($10^{-6}\,M_{\sun} \,yr^{-1}$) and $ r_{c}/R_{\ast}$, the location of 
      the critical point.}
         \label{tabla4}
     $$ 
	 \begin{array}{lccccccc}
            \hline \hline
            \noalign{\smallskip}
            & \Omega  & &V_{\infty} && F_{m} && r_{c}/R_{\ast}\\
            \noalign{\smallskip}
            \hline
            \noalign{\smallskip}
            pole    & 0.0&\,\,&1475&& 0.097 && 1.053 \\
 	    \hline
            equator & 0.7 & &  318 && 1.042 && 15.07 \\
            equator & 0.8 & &  292 && 1.087 && 16.04 \\    
            equator & 0.9 & &  267 && 1.138 && 18.41 \\
            equator & 0.99 & & 248 && 1.188 && 19.88 \\            %
	    \noalign{\smallskip}
            \hline
         \end{array}
     $$ 
   \end{table}

\section{Discussion    \label{diskusion}}
We want to stress the importance of the combined effect from slow and fast solutions
with bi--stability line--force parameters in forming an outflowing disk wind in B[e] 
supergiants. Density contrast of order of $10^2$ up to large distances from the star are 
attained.\\
This theoretical value is in qualitative agreement with the values derived from observations 
of order of $10^2$--$10^3$ (Zickgraf et al.~1989, Zickgraf 1998 and references therein, 
Bjorkman 1998).

Previous simulations of disk formation in rotating radiation driven winds induced by 
bi--stability have underestimated the density contrast, mainly due to: a) The use of 
a $\beta$--field (with $\beta=1$, Lamers \& Pauldrach~1991) to describe the wind velocity 
profile, even for high rotational speeds where this approximation fails (Cur\'e~2004). Table 
[1] from Lamers \& Pauldrach~(1991), shows that $\tau_{L} > 3$ exactly when the standard 
fast solution do not further exist, and b) Pelupessy et al.~(2000) calculations, based in 
the fast solutions, were restricted to values of $\Omega \le 0.6$ and these rotation values 
are not high enough for developing a dense disk. 

A dense disk is formed when the slow solution starts to exist. For our test star, this occurs 
for $\Omega \gtrsim 0.7$. This condition is in agreement with the estimation of $0.74 \le \Omega \le 0.79$ done by Zickgraf~(1998) in order to reproduce observable effects in the structure of 
stellar winds. However,  observational rotation speeds of B[e] supergiants have high uncentainties, 
because only a few stars show photospheric absorption lines appropiated for the measurements 
of $V\,\sin(i)$. The inferred observational value of $\Omega$ lies in the range $0.4$--$0.7$ (Zickgraf~1998).

We have verified that the $\delta$ line--force parameter, which is related  to changes in 
ionization troughout the wind, modifies the wind structure, mass--loss rate, terminal 
velocity and location of the singular point (compare Tables \ref{tabla3} and \ref{tabla4}). 
Despite these changes, the density ratio $\rho_{e}/\rho_{p}$ is almost the same as in the 
frozen--in ionization case.

Furthermore our result for the density contrast based on the m--CAK fast solution for 
$\Omega=0.6$ are of the same order as Pelupessy et al.~(2000), eventhough they included 
a more detailed description of the distorsion of the star due to the rotational speed. 

The existence of a fast solution for the polar direction and a slow solution for the 
equatorial direction, calculated using bi--stability line--force parameters, give a 
density contrast of order of $100$.

Since most of the B[e] supergiants in the H--R Diagram are located below the bi-stability 
jump temperature ($25,000$K), in our conception, the theoretical explanation for the 
existence of a two--component wind model (Zickgraf et al.~1985) is due to the nature 
of the solutions of rapidly rotating radiation driven wind. The change (jump) from fast 
solution to slow solution at some latitude accounts by itself a two--component wind, where 
each solution structure has its own set of line--force parameters. This picture would be 
remarked for cases when the bi--stability jump is present.

Another important aspect to remark is the scarcity of self--consistent calculations of 
line--force parameters $k$, $\alpha$, $\delta$ for the m--CAK fast solution and the lack 
of calculations for our slow solution. The uncertainty in the values of the parameters 
reflects in the value of the terminal velocity, mass loss rate, as well as in the density 
contrast. Specifically, the predicted terminal velocities, see tables \ref{tabla3} 
and \ref{tabla4}, are about $2$--$3$ times greater than values inferred by observations 
(Zickgraf 1998).

Therefore our results, that combine fast and slow wind solutions, are a first 
approximation to re-investigate disk formation in high rotating stars with radiation driven 
winds. A detailed wind model needs a self--consistent line--force parameter calculations 
for both, fast and slow solutions.

\section{Conclusions \label{conclu}}
We have revisited radiative driven wind models for a high rotating B[e] supergiant 
($\Omega \gtrsim 0.6$) assuming a change in the line--force parameters due to the 
bi--stability jump. The existence of slow and fast solutions in the model, predicts 
density contrast which are of order of $10^2 - 10^4$  near the stellar surface 
(r $\lesssim$  2 $R_{\ast}$), while outside it falls to values of about tens to hundreds 
and the disk extends up to $\sim 100$ stellar radii. 
Comparing the density contrast predicted by the CAK and m--CAK models, we conclude that 
the m--CAK model better describes disk formation in B[e] supergiants being in qualitative 
agreement with observations.

\begin{acknowledgements}
This work has been possible thanks to the research cooperation agreement UBA/UV, UNLP/UV 
and DIUV project 15/2003. 
\end{acknowledgements}

\end{document}